\documentclass[10pt,a4paper]{article}
\usepackage{geometry} % see geometry.pdf on how to lay out the page. There's lots.
\usepackage{amsmath}
\usepackage{amsthm}
\usepackage{amssymb}
\usepackage{amsfonts}
\usepackage[latin1]{inputenc}

\usepackage{cite}
\usepackage{soul}

\usepackage{color}

\def\be{\begin{equation}}
\def\ee{\end{equation}}
\def\bea{\begin{eqnarray}}
\def\eea{\end{eqnarray}}

\def\1{\'{\i}}                           
 
\def\conm#1#2{\left[ #1,#2 \right]}  
\def\pois#1#2{\left\{ #1,#2 \right\}}

 \def\ro{\eta}

\def\>#1{{\mathbf #1}}

\def\c{\Lambda}
\def\gc#1{\mathfrak{g}^{#1+1}_\Lambda}
\def\Gc#1{G^{#1+1}_\Lambda}

\def\kgc#1{(\mathfrak{g}^{#1+1}_\Lambda,\delta_\kappa)}
\def\kGc#1{\kappa \text{-} G^{#1+1}_\Lambda}
\def\Mc#1{\mathcal{M}^{#1+1}_\Lambda}

\begin{document}

 \begin{center}
\ 

\hfill

\bigskip

{\Large 
{\bf Quantum groups, non-commutative Lorentzian spacetimes  

\smallskip
and curved momentum spaces
  }}

\bigskip
\bigskip

{\sc I. Gutierrez-Sagredo$^\dag$\footnote{Based on the contribution presented at the ``First Hermann Minkowski Meeting on the Foundations of Spacetime Physics'' held in Albena, Bulgaria, May 15-18, 2017. \\
Proceedings of the First Hermann Minkowski Meeting on the Foundations of Spacetime Physics, Christopher Duston, Marc Holman (Editors).}, A. Ballesteros$^\dag$, G. Gubitosi$^{\dag}$, F.J. Herranz$^\dag$}

{$^\dag$ Departamento de F\1sica, Universidad de Burgos, 
E-09001 Burgos, Spain}

e-mail: {igsagredo@ubu.es, angelb@ubu.es, ggubitosi@ubu.es, fjherranz@ubu.es}

\end{center}

%\begin{center}\today\end{center}

\begin{abstract}
The essential features of a quantum group deformation of classical symmetries of General Relativity in the case with non-vanishing cosmological constant $\Lambda$ are presented. We fully describe (anti-)de Sitter non-commutative spacetimes and curved momentum spaces in $(1+1)$ and $(2+1)$ dimensions arising from the $\kappa$-deformed quantum group symmetries. These non-commutative spacetimes are introduced semiclassically by means of a canonical Poisson structure, the Sklyanin bracket, depending on the classical $r$-matrix defining the $\kappa$-deformation, while curved momentum spaces are defined as orbits generated by the $\kappa$-dual of the Hopf algebra of quantum symmetries. Throughout this construction we use kinematical coordinates, in terms of which the physical interpretation becomes more transparent, and the cosmological constant $\Lambda$ is included as an explicit parameter whose $\Lambda\to 0$ limit provides the Minkowskian case. The generalization of these results to the physically relevant $(3+1)$-dimensional deformation is also commented. 
\end{abstract}

%%%%%%%%%%%%%%%%%%%%%%%%%%%%%%%%%%%%%%%%%%%%%%%%%%%%%%%%%%

\tableofcontents

\contentsline {section}{Acknowledgments}{16}

\contentsline {section}{Appendix: Invariant vector fields}{16}

\contentsline {section}{References}{19}

%%%%%%%%%%%%%%%%%%%%%%%%%%%%%%%%%%%%%%%%%%%%%%%%%%%%%%%%%%

\section{Introduction}

Recent research in phenomenological aspects of quantum gravity suggests that Plank scale deformations of classical symmetries of General Relativity are to be expected when introducing quantum theory in the picture. An example is provided by Deformed Special Relativity theories (formerly introduced in~\cite{AmelinoCamelia:2000mn,AmelinoCamelia:2000ge,Amelino-Camelia:2002vy}  and further developed in~\cite{MagueijoSmolin, Kowalski-Glikman:2002we, Kowalskia, Kowalskib,  Lukierski:2002df, DSRjpa}) in which Planck energy, or equivalently Planck length, is introduced as a second relativistic invariant (besides the speed of light) which modifies the classical symmetries of the system. These kind of theories have attracted a considerable amount of attention during last years due to the observable phenomenology they could provide \cite{AmelinoCamelia:2008qg}. In this setting, quantum groups~\cite{Dri,CP, majid} (in which the deformation parameter is related with the Planck energy) are natural candidates to replace classical Lie groups of spacetime symmetries. These quantum deformations not only imply the emergence of non-commutative spacetimes (see for example~\cite{kMinkowski,Majid:1994cy,kZakr,LukR,LukNR,nullplaneR, RossanoPLB, BGHMN2017dice, ahep}) but also induce a non-trivial structure of momentum space~\cite{Majid:1999tc, AmelinoCamelia:1999pm, KowalskiGlikman:2002jr, Gubitosi:2013rna} (a long forgotten idea whose intuition is clearly stated in the so-called Born Reciprocity~\cite{BornReciprocity}). In relation with this non-trivial structure of momentum space, some concrete phenomenology has been described: firstly, curvature of momentum space induces a dual red-shift~\cite{Amelino-Camelia:2013uya}, i.e. an energy dependent distance covered in a given time by a free massless particle. Secondly, the so-called dual-gravity lensing~\cite{AmelinoCamelia:2011gy}, i.e. an energy dependent direction from which a particle emitted by a given source reaches a faraway detector.

The aim of this paper is, firstly, to review some of the key features of the $\kappa$-deformation for de Sitter and anti-de Sitter spaces, in such a way that the well-known results for the flat case are smoothly recovered by taking the limit $\c \rightarrow 0$. Secondly, we will summarize recent investigations in which momentum spaces arising from de Sitter, anti-de Sitter and Poincar\'e symmetries have been explicitly constructed~\cite{BGGHplb2017,BGGH31}. By presenting these two different features of the same quantum deformation in a unified way we attempt to clarify their relations and common origins.

In the first part of the paper we will present the coboundary Lie bialgebra structure which characterizes the $\kappa$-deformation, which is defined by a certain classical $r$-matrix. The cocommutator defined by this $r$-matrix endows the original Lie algebra with a Lie bialgebra structure, which is the tangent counterpart of the $\kappa$-deformed algebra and thus identifies the deformation. Such $\kappa$-Poisson-Hopf algebra  for the (1+1) and  (2+1)-dimensional cases (which is the semi-classical version of the full quantum algebra) will be explicitly presented.

At the Lie group level, the $\kappa$-Poisson-Lie structure on the  (anti-)de Sitter and Poincar\'e groups will be then given by the so-called Sklyanin bracket. By using exponential coordinates of the second kind on these Lie groups, we will give explicit expressions for the fundamental Poisson brackets of the subalgebra generated by the spacetime coordinates. We stress that the so-obtained non-commutative (Poisson homogeneous) spacetime is invariant under the Poisson-Lie group associated with the Poisson-Hopf algebra of deformed symmetries in the sense that the group action on the non-commutative Poisson spacetime defines a Poisson map. 

In the second part of the paper we will pay attention to the dual quantum group and its relation with the construction of momentum space. By computing the dual Poisson-Lie group to the Poisson-Hopf algebra of deformed spacetime symmetries, we will give a full description of the associated $\kappa$-deformed momentum space. Moreover, we will prove that a certain orbit of this dual group generates (half of) de Sitter space in $2 (n+1)-1 = 2n+1$ dimensions. The explicit expressions of this construction will be presented for the $(1+1)$ and $(2+1)$-dimensional cases ($n=1$ and $n=2$, respectively).

The central role played in this paper by the (2+1)-dimensional case is due to its relevance in Quantum Gravity, where it is often considered as a suitable toy model which incorporates some conceptual key points that a full quantum gravity theory is supposed to address (see \cite{Carlip2003book} for an excellent introduction to the topic). The fundamental reason for this is that (2+1)-dimensional general relativity is a topological theory, which in more physical terms means that there are no gravitational waves. This implies that all possible solutions of Einstein's field equations are locally isometric to the three model spacetimes (Minkowski, de Sitter or anti-de Sitter) depending only on the value of the cosmological constant. In this way, the study of these three possibilities gives a full understanding of (2+1) general relativity, modulo global topology. In fact, as proved in \cite{AT1986,Witten1988}, general relativity in (2+1) dimensions admits a reformulation as a Chern-Simons gauge theory, with the isometry group playing the role of the gauge group. This has motivated a lot of recent work regarding the compatibility of this Chern-Simons action with quantum group symmetries and their associated Poisson-Lie counterparts. For a detailed account of this subject, see \cite{MS2003poisson,BGHMN2017dice,MS2009generalized,BGH2018PoincareDD} and references therein.

The structure of the paper is the following. In section~\ref{Fundamental_concepts} we will summarize the (anti-)de Sitter and Poincar\'e classical symmetries of Lorentzian spacetimes, as well as the mathematical setting of quantum groups and the essential features of the  $\kappa$-deformation. Then we will describe in detail the $(1+1)$-dimensional case, including its non-commutative spacetime (section~\ref{kappa11}) and its momentum space (section~\ref{kappa11CMS}). In the same way, in sections~\ref{kappa21} and~\ref{kappa21CMS} their $(2+1)$-dimensional analogues will be presented. In the closing section some comments on the generalization of this construction to the physically relevant $(3+1)$-dimensional case will be discussed.

Some comments on the notation are in order. We will write $\gc n$ for the Lie algebra of isometries of the $(n+1)$-dimensional maximally symmetric spacetime $\Mc n$, which depending of the value of the cosmological constant $\c$ will be the so-called de Sitter $\c > 0$, Minkowski $\c = 0$ or anti-de Sitter $\c < 0$ spacetime. Their associated Lie group will be denoted by $\Gc n$, and by $\kgc n$ we will denote the Lie bialgebra associated to the $\kappa$-deformation of $\gc n$.

%%%%%%%%%%%%%%%%%%%%%%%%%%%%%%%%%%%%%%%%%%%%%%%%%%%%%%%%

\section{Fundamental concepts}\label{Fundamental_concepts}

In this section we give an introduction to the basic concepts and mathematical structures that will be used in the rest of the paper. Firstly, we  describe the spacetimes $\Mc n$ as coset spaces of its isometry groups by the Lorentz subgroup. Then, we give a motivating introduction to the mathematical concept of quantum group, which will be heavily used in the rest of the paper. Finally, we  introduce the $\kappa$-deformation, whose applications in the construction of non-commutative spacetimes and curved momentum spaces  in $(1+1)$ and $(2+1)$-dimensions are the main topic of this paper.

\subsection{Lorentzian kinematical groups in (1+1) and (2+1) dimensions}
The main purpose of this paper is the study of certain quantum deformations of Minkowski, de Sitter and anti-de Sitter spacetimes (the maximally symmetric Lorentzian spacetimes with constant curvature), namely the so-called $\kappa$-deformation, and their further consequences in momentum space. Moreover, our construction is based on a deformation of the group of isometries of these spacetimes, which will result in the construction of (the Poisson version of) the so-called quantum Poincar\'e, de Sitter and anti-de Sitter spacetimes. For the sake of completeness we shall review here some results and fix the notation used in the rest of the paper. 

Let $\c$ be the cosmological constant and let $\Mc n$ be the maximally symmetric Lorentzian spacetime with curvature   proportional to $\c$ with $n$ space dimensions (and one temporal dimension). Then denote its isometry group (its identity component to be rigorous)
by $\Gc n$. The Lie algebra of this group will be denoted by $\gc n$. From this point of view maximally symmetric spacetimes are defined as cosets $\Gc n /H^{n+1}$, where $H^{n+1}$ is the Lorentz group (the stabilizer of the origin). 

In $(1+1)$ dimensions we have
 \[
\gc 1 = 
\left \{
  \begin{tabular}{ccc}
  $\mathfrak{so}(1,2)$ & if & $\c < 0$ \\
  $\mathfrak{iso}(1,1)$  & if & $\c = 0$ \\
  $\mathfrak{so}(2,1)$  & if & $\c > 0$ 
  \end{tabular}
\right. , \qquad
 \Gc 1 = 
\left \{
  \begin{tabular}{ccc}
  SO(1,2) & if & $\c < 0$ \\
  ISO(1,1)  & if & $\c = 0$ \\
  SO(2,1)  & if & $\c > 0$ 
  \end{tabular}
\right. ,
\]
 \[
\Mc 1 = 
\left \{
  \begin{tabular}{ccc}
  SO(1,2)/SO(1,1) & if & $\c < 0$ \\
  ISO(1,1)/SO(1,1)  & if & $\c = 0$ \\
  SO(2,1)/SO(1,1)  & if & $\c > 0$ 
  \end{tabular}
\right. .
\]

The $\gc 1$ algebra written in the kinematical basis $\{P_0,P_1,K\}$ is defined by the brackets
\be
\conm{K}{P_0}=P_1,
\qquad
\conm{K}{P_1}= P_0 ,
\qquad
\conm{P_0}{P_1}=-\c \,K,
\label{ds11}
\ee
where $K$ is the generator of boost transformations, $P_0$ and $P_1$ are the time and space translation generators. Note that in the $(1+1)$-dimensional case the de Sitter ($\c > 0$) and anti-de Sitter ($\c < 0$) algebras and groups are isomorphic; this is no longer true in higher dimensions.

In $(2+1)$ dimensions we have
 \[
\gc 2 = 
\left \{
  \begin{tabular}{ccc}
  $\mathfrak{so}(2,2)$ & if & $\c < 0$ \\
  $\mathfrak{iso}(2,1)$  & if & $\c = 0$ \\
  $\mathfrak{so}(3,1)$  & if & $\c > 0$ 
  \end{tabular}
\right. ,\qquad 
 \Gc 2 = 
\left \{
  \begin{tabular}{ccc}
  SO(2,2) & if & $\c < 0$ \\
  ISO(2,1)  & if & $\c = 0$ \\
  SO(3,1)  & if & $\c > 0$ 
  \end{tabular}
\right. ,
\]
\[
\Mc 2 = 
\left \{
  \begin{tabular}{ccc}
  SO(2,2)/SO(2,1) & if & $\c < 0$ \\
  ISO(2,1)/SO(2,1)  & if & $\c = 0$ \\
  SO(3,1)/SO(2,1)  & if & $\c > 0$ 
  \end{tabular}
\right. .
\]
  
The $\gc 2$ algebra in the kinematical basis $\{P_0,P_1,P_2,K_1,K_2,J\}$ is defined by the brackets
\be
\begin{array}{lll} 
\conm{J}{P_i}=   \epsilon_{ij}P_j , &\qquad
\conm{J}{K_i}=   \epsilon_{ij}K_j , &\qquad  \conm{J}{P_0}= 0  , \\[2pt]
\conm{P_i}{K_j}=-\delta_{ij}P_0 ,&\qquad \conm{P_0}{K_i}=-P_i ,&\qquad
\conm{K_1}{K_2}= -J   , \\[2pt]
\conm{P_0}{P_i}=-\Lambda\, K_i ,&\qquad \conm{P_1}{P_2}= \Lambda\, J  ,
\end{array}
\label{ds21} 
\ee
where $i,j \in \{1,2\}$, and  $\epsilon_{ij}$ is the skew-symmetric tensor with $\epsilon_{12}=1$. Here $J$ is the generator of spatial rotations, $K_1,K_2$ are generators of boosts and $P_0,P_1,P_2$ are generators of temporal and spatial translations, respectively. Note that under the projection onto the subspace spanned by $\{P_0,P_1,K_1\equiv K \}$ the Lie algebra $\gc 1$ (\ref{ds11})  is recovered. 

%%%%%%%%%%%%%%%%%%%%%%%%%%%%%%%%%%%%%%%%%%%%%%%%%%%%%%%%

\subsection{Quantum groups and Lie bialgebras}

Quantum groups are quantizations of  Poisson-Lie  groups, {\em i.e.}, quantizations of the Poisson-Hopf algebras of multiplicative Poisson structures on Lie groups~\cite{Dri,CP, majid}. It is well known that Poisson-Lie structures on a (connected and simply connected) Lie group $G$ are in one-to-one correspondence with Lie bialgebra structures $(\mathfrak{g},\delta)$ on $\mathfrak{g}=\text{Lie}(G)$~\cite{DriPL}, where the skewsymmetric cocommutator map $\delta:\mathfrak{g} \to \mathfrak{g}\wedge \mathfrak{g}$
fulfils two conditions:
\begin{itemize}
\item i) $\delta$ is a 1-cocycle, {\em  i.e.},
\be
\delta([X,Y])=[\delta(X),\,  Y\otimes 1+ 1\otimes Y] + 
[ X\otimes 1+1\otimes X,\, \delta(Y)] ,\qquad \forall \,X,Y\in
\mathfrak{g}.
\label{1cocycle}\nonumber
\ee
\item ii) The dual map $\delta^\ast:\mathfrak{g}^\ast\wedge \mathfrak{g}^\ast \to \mathfrak{g}^\ast$ is a Lie bracket on $\mathfrak{g}^\ast$.
\end{itemize}
Therefore, each quantum group $G_z$ (with quantum deformation parameter $z= \ln q$) can be associated with a Poisson-Lie group $G$, and the latter with a unique Lie bialgebra structure $(\mathfrak{g},\delta)$.

On the other hand,  the dual version of quantum groups are quantum algebras ${\mathcal U}_z(\mathfrak{g})$, which are Hopf algebra deformations of universal enveloping algebras ${\mathcal U}(\mathfrak{g})$, and are constructed as formal power series in the deformation parameter $z$
and coefficients in ${\mathcal U}(\mathfrak{g})$. The Hopf algebra structure in ${\mathcal U}_z(\mathfrak{g})$ is provided by a coassociative coproduct map
$
\Delta_z: {\mathcal U}_z(\mathfrak{g})\longrightarrow {\mathcal U}_z(\mathfrak{g})\otimes {\mathcal U}_z(\mathfrak{g})
$,
which is an algebra homomorphism, together with its associated counit $\epsilon$ and antipode $\gamma$ mappings. If we write the coproduct as a formal power series of maps, namely
\be
\Delta_z
=\Delta_0 + z\,\delta+ o[z^2] ,
\label{powerco}\nonumber
\ee
then the skew-symmetric part of the first-order deformation (in $z$) of the coproduct map
is just the Lie bialgebra cocommutator map $\delta$, where we have denoted the primitive (undeformed) coproduct for ${\mathcal U}(\mathfrak{g})$ as $\Delta_0(X)=X \otimes 1+1\otimes X $. 

In this way, each quantum deformation becomes related to a unique Lie bialgebra structure $(\mathfrak{g},\delta)$. 
More explicitly, if we consider a basis for $\mathfrak{g}$ where
\be
[X_i,X_j]=c^k_{ij}X_k ,
\label{liealg}\nonumber
\ee
any cocommutator $\delta$ will be of the form
\be
\delta(X_i)=f^{jk}_i\,X_j\wedge X_k \, ,
\label{precoco}\nonumber
\ee
where $f^{jk}_i$ is the structure tensor of the dual Lie algebra $\mathfrak{g}^\ast$, that will be given by
\be
[\hat\xi^j,\hat\xi^k]=f^{jk}_i\,\hat\xi^i \, ,
\label{dualL}
\ee
where $\langle  \hat\xi^j,X_k \rangle=\delta_k^j$. Notice that the cocycle condition for the cocommutator $\delta$   implies the following compatibility equations among the structure constants $c_{ij}^k$ and $f_k^{ij}$:
$$
f^{ab}_k c^k_{ij} = f^{ak}_i c^b_{kj}+f^{kb}_i c^a_{kj}
+f^{ak}_j c^b_{ik} +f^{kb}_j c^a_{ik}. 
$$

The connection of these structures with non-commutative spacetimes arises when $G$ is a group of isometries of a given spacetime (for instance $\Gc n$ for the spacetime $\Mc n$). Then $X_i$ will be the Lie algebra generators and $\hat\xi^j$ will be the local coordinates on the group. If we have a non-trivial deformation of $\mathcal U (\mathfrak{g})$, then the cocommutator $\delta$ is non-vanishing and the commutator~\eqref{dualL} among the spacetime coordinates associated to the translation generators of the group will be non-zero. This is  just the way in which non-commutative spacetimes arise from quantum groups. Moreover, higher-order contributions to the non-commutative spacetime~\eqref{dualL} can be obtained from higher orders of the full quantum coproduct  $\Delta_z$. 

It is worth mentioning that in many cases the 1-cocycle $\delta$ is found to be a coboundary 
\be
\delta(X)=[ X \otimes 1+1\otimes X ,\,  r],\qquad 
\forall\,X\in \mathfrak{g} ,
\label{cocom}
\ee
where the classical $r$-matrix 
$
r=r^{ij}\,X_i \wedge X_j\, ,
$
is a solution of the modified classical Yang--Baxter equation  
\be
[X\otimes 1\otimes 1 + 1\otimes X\otimes 1 +
1\otimes 1\otimes X,[[r,r]]\, ]=0, \qquad \forall X\in \mathfrak{g},
\label{mCYBE}\nonumber
\ee
and the Schouten bracket is defined as
$$
[[r,r]]:=[r_{12},r_{13}]+ [r_{12},r_{23}]+ [r_{13},r_{23}]  ,
$$
  where $r_{12}=r^{ij}\,X_i \otimes X_j\otimes 1, \, r_{13}=r^{ij}\,X_i \otimes 1\otimes X_j, \, r_{23}=r^{ij}\,1 \otimes X_i\otimes X_j $. Recall that $[[r,r]]=0$ is just the classical Yang--Baxter equation. In these coboundary cases the $r$-matrix identifies both the Lie bialgebra and the quantum deformation completely.

  %%%%%%%%%%%%%%%%%%%%%%%%%%%%%%%%%%%%%%%%

\subsection{The $\kappa$-deformation of Lorentzian kinematical groups}

In the rest of this paper we will concentrate on spacetime and momentum space deformations induced by a certain well-known deformation of Lorentzian symmetries, the so-called $\kappa$-deformation. The $\kappa$-deformation was originally introduced in~\cite{LukierskiRuegg1992, Giller,Lukierskibc} for the case of vanishing cosmological constant and the associated non-commutative spacetime was also constructed~\cite{kMinkowski} (see also~\cite{LukNR}). This non-commutative spacetime has as its main feature the non-commutativity between space $\hat x^i$ and time $\hat x^0$  coordinates, whereas space coordinates commute:
$$
[ \hat x^i,\hat x^0] = \frac{1}{\kappa}\,  \hat x^i ,\qquad [ \hat x^i,\hat x^j] = 0,\qquad \Lambda=0 .
$$
Thus  this non-commutativity is controlled by a parameter called $\kappa$ (in terms of the usual  deformation parameter $q=\ln (1/\kappa$)), which is somehow related with the Planck length or Planck energy, in such a way that the limit $\kappa \rightarrow \infty$ recovers the classical Minkowski spacetime and the Poincar\'e group. 

Some years later, this construction was generalized to the case of non-vanishing cosmological constant in $(1+1)$ and $(2+1)$ dimensions,  giving rise to the so-called $\kappa$-de Sitter and $\kappa$-anti-de Sitter groups~\cite{ck2, ck3, starodutsev} and their associated non-commutative spacetimes~\cite{BGHMN2017dice,ahep}. Only recently~\cite{BHMNplb2017} the Poisson version of the $\kappa$-(anti-)de Sitter  algebra in $(3+1)$ dimensions has been constructed. It has been made clear in these studies that the interplay between the cosmological constant $\c$ and the deformation parameter $z\equiv 1/\kappa$\, is quite intricate and the explicit construction of deformed symmetries in $(3+1)$ dimensions is highly non-trivial.~\footnote{In our notation $\kappa = 1/z$, so that  the commutative limit is $z \rightarrow 0$. The notation $\kappa$-deformation is here maintained due  to historical reasons, although we could as well speak about $z$-deformation.} Moreover, the associated (3+1) non-commutative spacetime is still work in progress~\cite{BGH31}.

Mathematically, the $\kappa$-deformation can be defined as a Hopf algebra structure on $\mathcal{U}_z (\gc n)$ in the direction of the coboundary Lie bialgebra $\kgc n$, defined by an $r$-matrix  that we will call $r^{n+1}$. The explicit form of this $r$-matrix is dimension dependent, but the structure is always similar: it contains products of boost and translation generators as well as products of rotations weighted by the square root of the cosmological constant. In the case of vanishing cosmological constant it can be written in our kinematical basis as 
\be \label{rkappamink}
r^{n+1}_{0}=z \sum_{i=1}^n K_i \wedge P_i\,.
\ee
In the case of non-vanishing cosmological constant $\c$ this $r$-matrix is not modified when $n=1$ or $n=2$, but in the $(3+1)$-dimensional case a new term must be added~\cite{BGGH31,BHMNplb2017, LBC, prague}, resulting in 
\be \label{r31ads}
r^{3+1}_\c=z \sum_{i=1}^3 K_i \wedge P_i + z\,\sqrt{-\c}\, J_1 \wedge J_2\,.
\ee
This last term implies a deformation of the three-dimensional rotation subalgebra, which cannot be avoided by any generalization of the $\kappa$-Poincar\'e algebra (see~\cite{BGH31} for a complete discussion).

%%%%%%%%%%%%%%%%%%%%%%%%%%%%%%%%%%%%%%%%%%%%%%%%%%%%%%%%%%

\section{The $\kappa$-$G_\Lambda$ quantum group in (1+1) dimensions}\label{kappa11}

In this section we will present the Poisson-Hopf algebra structure on the universal enveloping algebra $\mathcal{U}_z(\gc 1)$ induced by the $\kappa$-deformation. Firstly we give the full description of the Lie bialgebra $\kgc 1$ and then we present the explicit expressions for the Poisson-Hopf structure on $\mathcal{U}_z(\gc 1)$. After that, we consider the Lie group $\Gc 1$ and a concrete Poisson structure, the so-called Sklyanin bracket, which endows $\kGc 1$ with a Poisson-Lie structure. Explicit expressions for the (Poisson version of the) non-commutative spacetime are given. We conclude by performing the limit $\c \rightarrow 0$ and recovering well-known expressions for $\kappa$-Minkowski spacetime.

%%%%%%%%%%%%%%%%%%%%%%%%%%%%%%%%%%%%%%%%%%%%%%%%%%%%%%%%%%

\subsection{Coboundary Lie bialgebra structure $\kgc 1$}

The Lie algebra brackets for $\gc 1$ are  given in the kinematical basis $\{P_0,P_1,K\equiv K_1\}$ in (\ref{ds11})   and the $r$-matrix defining the Lie bialgebra $\kgc 1$  reads~\cite{LBC,PL2}
$$
r^{1+1}_\c=z K \wedge P_1\, .
$$
The associated cocommutator can be directly obtained  from~\eqref{cocom} and is given by
\be
\delta_\kappa(P_0)=0, \qquad\ 
\delta_\kappa(P_1)= z\, P_1 \wedge P_0, \qquad
\delta_\kappa(K)=z\, K \wedge P_0.
\label{delta11}
\ee
The quadratic Casimir for $\gc 1$ turns out to be
\be
{\mathcal C}=P_0^2-P_1^2+\c \, K^2\,.
\label{cas11}
\ee

  %%%%%%%%%%%%%%%%%%%%%%%%%%%%%%%%%%%%%%%%%%%%%%%%%%%%%%%%%%

\subsection{Poisson-Hopf algebra structure on $\mathcal{U}_z(\gc 1)$}
  
  The Poisson version of $\gc 1$ (\ref{ds11})    is defined by the Poisson brackets
$$
\pois{K}{P_0}=P_1,
\qquad
\pois{K}{P_1}= P_0 ,
\qquad
\pois{P_0}{P_1}=-\c \,K,
$$
and the undeformed Poisson-Hopf structure is given by the primitive coproduct 
$$
\Delta_0 (X)=X \otimes 1 + 1 \otimes X , \qquad \forall X \in \gc 1.
$$

To make contact with previous results~\cite{Kowalski-Glikman:2013rxa,BGGHplb2017,BGGH31} we will work from now on in the so-called bicrossproduct-type basis~\cite{Majid:1994cy}, which is related to the previous basis by the following nonlinear redefinition of the generators
$$
P_0 \to P_0,
\qquad
P_1 \to e^{\frac{z}{2} P_0} \,P_1,
\qquad
K \to e^{\frac{z}{2} P_0} \,K,
$$
so that the deformed Poisson-Hopf algebra becomes
\be
\pois{K}{P_0}=P_1,
\qquad
\pois{K}{P_1}=\frac{ 1-\exp(-2 z P_0)}{2z}-\frac{z}{2}\,(P_1^2 - \c\, K^2) ,
\qquad
\pois{P_0}{P_1}= -\c \,K,
\label{kdsa11}
\ee
with deformed coproduct  
\begin{eqnarray}
&& \Delta_\kappa(P_0) = P_0 \otimes 1 + 1 \otimes P_0, \nonumber\\
&&\Delta_\kappa(P_1)=P_1\otimes 1 +  e^{-{z}P_0} \otimes P_1,\label{cop11}\\
&& \Delta_\kappa(K)= K\otimes 1 +  e^{-{z} P_0} \otimes K.
\nonumber
\end{eqnarray}  
It is worth noting here that a deformation at the coalgebra level (coproduct) implies a specific deformation at the Poisson algebra level (Poisson brackets), since the former has to be a Poisson algebra homomorphism with respect to the latter. This deformation also affects the quadratic Casimir~\eqref{cas11}, resulting in the modified Casimir 
\be
{\mathcal C}_z=\left(\frac{\sinh\left( z P_0 /2 \right)}{z/2}\right)^2 - e^{{z} P_0} (P_1^2 - \c \, K^2) \, ,
\label{kcas11}
\ee
which could be interpreted as a deformed dispersion relation~\cite{BGGHplb2017}. A clearer intuition for this interpretation will be given when constructing associated momentum spaces in the next section.

%%%%%%%%%%%%%%%%%%%%%%%%%%%%%%%%%%%%%%%%%%%%%%%%%%%%%%%%

\subsection{Non-commutative spacetime induced by a Poisson-Lie structure on $\Gc 1$}

Until now we have looked at the consequences of the $\kappa$-deformation for spacetime symmetries, but these deformations also affect spacetime itself (later on we will see that also momentum space is affected, so we can say that these kind of symmetry deformations lead to a whole deformed phase space). The approach followed here to study (the Poisson versions of) non-commutative spacetimes heavily relies on the following two facts:
\begin{itemize}
\item The $\kappa$-deformation is defined by means of a coboundary Lie bialgebra;
\item Any coboundary Lie bialgebra $(\mathfrak{g}=\text{Lie}(G),\delta)$ canonically defines a Poisson-Lie structure on $G$.
\end{itemize}
This canonical Poisson-Lie structure on $G$ is the so-called Sklyanin bracket~\cite{CP}, given by
\be
\{f,g\} = r^{ij} \left(  X_i^L f X_j^L g - X_i^R f X_j^R g \right),
\label{skb}
\ee
in terms of the components $r^{ij}$ of the classical $r$-matrix defining the $\kappa$-deformation, and the  left- and right-invariant vector fields $X_i^L,X_i^R$ on the Lie group $G$. 

For the $(1+1)$-dimensional case under consideration, the Appendix contains the construction of  $\Gc 1$ using easy-to-interpret coordinates and explicit expressions for $X_i^L,X_i^R$, which are displayed in  Table \ref{IVF11}. Using these results, we can write an explicit expression for the (Poisson version of the) non-commutative spacetime just as the fundamental Poisson brackets associated to time and space coordinates, which read
\be
\pois{x^1}{x^0}=z \frac{\tanh{(\sqrt{-\c} x^1)}}{\sqrt{-\c}}=z \frac{\tan{(\sqrt{\c} x^1)}}{\sqrt{\c}}\,.
\label{nc11}
\ee
As we have emphasized above, this expression is the Poisson version of the full non-commutative spacetime with non-vanishing cosmological constant, which is not easy to be constructed by making use of Hopf algebra duality from the quantum algebra due to reordering contributions. However, in this particular case no ordering problems appear in~\eqref{nc11} and the full non-commutative spacetime is given by
$$
\conm{\hat x^1}{\hat x^0}=z \frac{\tanh{(\sqrt{-\c} \hat x^1)}}{\sqrt{-\c}}=z \frac{\tan{(\sqrt{\c} \hat x^1)}}{\sqrt{\c}}\,,
$$
where now $\hat x^0,\hat x^1$ are to be understood as operators acting on some suitable space of fundamental states.

%%%%%%%%%%%%%%%%%%%%%%%%%%%%%%%%%%%%%%%%%%%%%%%%%%%%%%%%

\subsection{Poincar\'e-Minkowski limit: contraction $\c\to 0$}

To finish with our study of spacetime consequences of the $\kappa$-deformation in $(1+1)$ dimensions, we will explicitly show how in the limit $\c \to 0$ we recover the results for the well-known $\kappa$-Poincar\'e case. Note that the coproduct is not modified in this case (the cosmological constant only appears in the Lie algebra and the coalgebra structure is independent of $\c$). The Poisson-Hopf structure in the limit $\Lambda\to 0$ is given by
\be
\pois{K}{P_0}=P_1,
\qquad
\pois{K}{P_1}=\frac{ 1-\exp(-2 z P_0)}{2z}-\frac{z}{2}\,P_1^2,
\qquad
\pois{P_0}{P_1}= 0,
\label{kdsa11b}
\ee
and the associated coproduct map by
\begin{eqnarray}
&& \Delta_\kappa(P_0) = P_0 \otimes 1 + 1 \otimes P_0, \nonumber\\
&&\Delta_\kappa(P_1)= P_1\otimes 1 +  e^{-{z}P_0} \otimes P_1,\label{cop11mink}\\
&& \Delta_\kappa(K)= K\otimes 1 +  e^{-{z} P_0} \otimes K.
\nonumber
\end{eqnarray}  
The deformed quadratic Casimir reads
\be
{\mathcal C}_z=\left(\frac{\sinh\left( z P_0 /2 \right)}{z/2}\right)^2 - e^{{z} P_0}P_1^2.
\label{kcas11b}
\ee
Finally, the Poisson non-commutative spacetime is given by
$$
\pois{x^1}{x^0}=z\; x^1
$$
that can be trivially quantized giving the full non-commutative spacetime
$$
\conm{\hat x^1}{\hat x^0}=z\; \hat x^1,
$$
which is the well-known $\kappa$-Minkowski spacetime introduced in~\cite{kMinkowski}.

%%%%%%%%%%%%%%%%%%%%%%%%%%%%%%%%%%%%%%%%%%%%%%%%%%%%%%%%%%

\section{Curved momentum spaces in (1+1) dimensions\label{kappa11CMS}}

In this section we review some recent results concerning the effects of the presence of a cosmological constant $\c$ on the momentum space of the $\kappa$-deformation. The fact that a quantum deformation could imply a non-trivial structure of momentum space was first proposed in~\cite{Snyder}. This is related with Born's Reciprocity idea~\cite{BornReciprocity}. Here we use the language of Hopf algebras and quantum groups, using the quantum duality principle as the key ingredient to construct a deformed momentum space. In fact the first explicit construction of a curved momentum space using this language was that of the associated momentum space to $\kappa$-Poincar\'e deformation, and this construction was reviewed in~\cite{Kowalski-Glikman:2013rxa}. The presence of spacetime curvature induced by $\c$ in addition to the quantum deformation has been a long standing problem obstructing the construction of the momentum space associated to $\kappa$-(anti-)de Sitter, and this issue has been recently solved in~\cite{BGGHplb2017}. We will see how the presence of $\c$ enforces us to enlarge the momentum space by including boost generators on an equal footing as translation generators. The need of this enlarged space will become evident once the underlying Lie bialgebra is considered, since in the latter spatial translation generators and boost generators play algebraically equivalent roles, as shown in~\cite{BGGH31}.

%%%%%%%%%%%%%%%%%%%%%%%%%%%%%%%%%%%%%%%%%%%%%%%%%%%%%%%%

\subsection{Dual Poisson-Lie group}

Consider the Lie bialgebra $\kgc 1$ whose cocommutator is given by~\eqref{delta11}. If we denote by  $\{X^0,X^1,L\}$  the  generators dual to, respectively, $\{P_0,P_1,K\}$, the dual Lie algebra $(\gc 1)^\ast$ is  given by the Lie brackets
\be
\conm{X^0}{X^1}=-z\,X^1,
\qquad
\conm{X^0}{L}=-z\,L,
\qquad
\conm{X^1}{L}=0.
\nonumber
\ee
Note that the quantum parameter $z$ that controls the quantum deformation appears in the above  Lie bracket of the dual algebra, while it only appears in the coalgebra structure of the original bialgebra (\ref{delta11}). A faithful representation $\rho : (\gc 1)^\ast \rightarrow \text{End}(\mathbb{R}^4)$ of this dual Lie algebra is given by 
\be
\rho(X^0)=z \left( 
\begin{array}{cccc}
0 & 0 & 0 & 1\\
0 & 0 & 0 & 0\\
0 & 0 & 0 & 0\\
1 & 0 & 0 & 0
\end{array}
\right) , \quad 
\rho(X^1)=z\left( 
\begin{array}{cccc}
0 & 1 & 0 & 0\\
1 & 0 & 0 & 1\\
0 & 0 & 0 & 0\\
0 & -1 & 0 & 0
\end{array}
\right)  ,\quad 
\rho(L)=z \sqrt{\Lambda}\left( 
\begin{array}{cccc}
0 & 0 &1 & 0\\
0 & 0 & 0 & 0\\
1 & 0 & 0 & 1\\
0 & 0 & -1 & 0
\end{array}
\right).    
\nonumber
\ee
This representation is faithful if and only if $\c \neq 0$. In fact, as $\sqrt{\c}$ appears, to avoid complex quantities we will assume from now on that $\c > 0$, that is, the de Sitter case. It should be noted that for anti-de Sitter with $\c < 0$ a completely analogous construction is possible and results are completely equivalent (see final comments in~\cite{BGGHplb2017}).

If we introduce exponential coordinates of the second kind $\{p_0,p_1,\chi \}$ on the dual Lie group, a group element $h$ sufficiently close to the identity  is given by
\be
h=\exp \left(p_1 \rho(X^1) \right) \exp \left( \chi\rho(L) \right) \exp \left(p_0 \rho(X^0) \right).
\nonumber
\ee
A straightforward computation leads to the following explicit parametrization 
\be
h=
\left( 
\begin{array}{cccc}
\cosh(z p_0) \, +\frac{1}{2}\,z^2\,e^{z\,p_0}\, (p_1^2 + \Lambda \chi^2)& z p_1 & z \sqrt{ \Lambda} \, \chi & \sinh(z p_0) \, +\frac{1}{2}\,z^2\,e^{z\,p_0}\, (p_1^2 + \Lambda \chi^2)\\
z\,e^{z\,p_0}\, p_1& 1 & 0 &z\, e^{z\,p_0}\, p_1\\
z\, e^{z\,p_0}\,  \sqrt{\Lambda} \, \chi & 0 & 1 &z\, e^{z\,p_0}\,  \sqrt{\Lambda}\, \chi\\
\sinh(z p_0) \, -\frac{1}{2}\, z^2\,e^{z\,p_0}\, (p_1^2 +\Lambda \chi^2) & -z p_{1} & -z \sqrt{\Lambda} \, \chi & \cosh(z p_0) \, -\frac{1}{2}\, z^2\,e^{z\,p_0}\, (p_1^2 + \Lambda \chi^2)
\end{array}
\right).
\nonumber
\ee
We denote this dual Lie group by $(\Gc 1)^\ast$ and we have that $(\gc 1)^\ast=\text{Lie}((\Gc 1)^\ast)$.

The multiplication law for the group $(\Gc 1)^\ast$ can be written as a co-product (see~\cite{dualJPA}) in the form
\be
\Delta(p_0)=p_0 \otimes 1 + 1 \otimes p_0, \qquad
\Delta(p_1)=p_1\otimes 1 +  e^{-{z} p_0} \otimes p_1, \qquad
\Delta(\chi)= \chi\otimes 1 +  e^{-{z} p_0} \otimes \chi.
\label{codual}
\ee
Following the quantum duality principle, this coproduct is just~\eqref{cop11} for $\mathcal{U}_z(\gc 1)$ if dual group coordinates and the generators of the Poisson-Hopf algebra $\mathcal{U}_z(\gc 1)$ are identified as follows:
\be
p_0\equiv P_0, \quad
p_1\equiv P_1, \quad
\chi\equiv K.
\label{id11}
\ee

Moreover, by following the technique presented in~\cite{dualJPA} it can be shown that the unique Poisson-Lie structure on $(\Gc 1)^\ast$ that is compatible with the coproduct~\eqref{codual} and has the Lie algebra $\gc 1$~\eqref{ds11} as its linearization is given by the Poisson brackets
\be
\pois{\chi}{p_0}=p_1,
\qquad
\pois{\chi}{p_1}=\frac{ 1-\exp(-2 z p_0)}{2z}-\frac{z}{2}\,(p_1^2  - \Lambda\, \chi^2) ,
\qquad
\pois{p_0}{p_1}=- \Lambda\,\chi,
\nonumber
\ee
which is exactly the Poisson-Hopf algebra $\mathcal{U}_z(\gc 1)$~\eqref{kdsa11} under the identification~\eqref{id11}.
Evidently, the Casimir function for this Poisson bracket is
\be
\mathcal{C}_z=\left(\frac{\sinh\left( z p_0 /2 \right)}{z/2}\right)^2 - e^{{z} p_0} (p_1^2  - \Lambda \chi^2).
\label{kcas11dual}
\ee

In this way, the composition law for the momenta with $\kappa$-de Sitter symmetry~\eqref{cop11} has been reobtained as the group law~\eqref{codual} for the coordinates of the dual Poisson-Lie group $(\Gc 1)^\ast$, and the $\kappa$-de Sitter Casimir function~\eqref{kcas11} can be interpreted as an on-shell relation~\eqref{kcas11dual} for these coordinates. 

As said before the main novelty with respect to the $\kappa$-Poincar\'e case described in~\cite{Kowalski-Glikman:2013rxa} is that here $(\Gc 1)^\ast$ is three-dimensional, and the momentum space associated to the $\kappa$-deformation is parametrized by $\{p_0,p_1,\chi \}$ and not only by the  momenta associated to spacetime translations. Moreover, both in the coproduct~\eqref{codual} and the Casimir function~\eqref{kcas11dual} the role of the parameters $\chi$ and $p_1$ is identical, which supports the role of the former as an additional ``hyperbolic" momentum for quantum symmetries with non-vanishing cosmological constant $\c$. As stated before this construction has been explicitly done for $\c > 0$ but an analogue for $\c < 0$ can be constructed by following the very same procedure.

%%%%%%%%%%%%%%%%%%%%%%%%%%%%%%%%%%%%%%%%%%%%%%%%%%%%%%%%

\subsection{Momentum space as an orbit and deformed dispersion relation}

In order to give a geometric interpretation of this enlarged momentum space, we follow the procedure formerly proposed in~\cite{Kowalski-Glikman:2013rxa} for vanishing cosmological constant $\c = 0$ and recently generalized in~\cite{BGGHplb2017} for $\c \neq 0$. Consider the action $(\Gc 1)^\ast \triangleright \mathbb{R}^{1,3}$ given by left multiplication, the point $(0,0,0,1) \in \mathbb{R}^{1,3}$ and the orbit of $(\Gc 1)^\ast$ passing through it. We can easily describe this orbit in terms of Cartesian coordinates $S_i$ on $\mathbb{R}^{1,3}$ as the set
\be
\left\{\left(S_0,S_1,S_2,S_3 \right) \in \mathbb{R}^{1,3} : -S_0^2 + S_1^2 + S_2^2 + S_3^2 =1,\ S_0+S_3>0\right\}, \label{ms11set}
\ee
which is nothing but a subset of de Sitter spacetime in $(2+1)$ dimensions written as the usual embedding in $\mathbb{R}^{1,3}$. Moreover, we can use previous coordinates $\{p_0,p_1,\chi \}$ on the Lie group $(\Gc 1)^\ast$ to parametrize this space. Explicitly we can write 
\begin{eqnarray}
&& S_0=\sinh(z p_0) \, +\frac{1}{2}\,z^2\,e^{z\,p_0}\, (p_1^2 + \Lambda \chi^2), \nonumber\\
&&  S_1=z\,  e^{z\,p_0}\, p_1, \nonumber\\
&& S_2= z\,e^{z\,p_0}\,  \sqrt{\Lambda}\, \chi,   \nonumber \\
&& S_3 = \cosh(z p_0) \, -\frac{1}{2}\,z^2\,e^{z\,p_0}\, (p_1^2 + \Lambda \chi^2). \nonumber
\end{eqnarray}  
The fact that this orbit is not the whole de Sitter spacetime in $(2+1)$ dimensions in encoded in the condition 
\be
S_0+S_3=e^{z\,p_0}>0,
\nonumber
\ee
so we can say that this orbit is half of $(2+1)$-dimensional de Sitter spacetime. Note that the orbit passing through the point $(0,0,0,\alpha)$, with $\alpha\neq 0$, would satisfy  $-S_0^2 + S_1^2 + S_2^2 + S_3^2 =\alpha^2$, so the resulting de Sitter spacetime radius is free.

%%%%%%%%%%%%%%%%%%%%%%%%%%%%%%%%%%%%%%%%%%%%%%%%%%%%%%%%

\subsection{Poincar\'e-Minkowski $\c \to 0$ limit  in momentum space}

We finish the study of the momentum space associated to the $\kappa$-deformation in $(1+1)$ dimensions by giving explicitly  the Poisson-Hopf structure on the dual group $(\Gc 1)^\ast$ in the limit $\c \to 0$ and the geometric interpretation of the orbit, obtaining in this way the known results for the $\kappa$-Minkowski momentum space~\cite{Kowalski-Glikman:2013rxa}.

The fundamental Poisson brackets are given by
\be
\pois{\chi}{p_0}=p_1,
\qquad
\pois{\chi}{p_1}=\frac{ 1-\exp(-2 z p_0)}{2z}-\frac{z}{2}\,p_1^2
 ,\qquad
\pois{p_0}{p_1}=0,
\nonumber
\ee
and the coproduct by
\be
\Delta(p_0)=p_0 \otimes 1 + 1 \otimes p_0, \qquad
\Delta(p_1)=p_1\otimes 1 +  e^{-{z} p_0} \otimes p_1, \qquad
\Delta(\chi)= \chi\otimes 1 +  e^{-{z} p_0} \otimes \chi.
\nonumber
\ee
The Casimir function for this Poisson structure reads
\be
\mathcal{C}_z=\left(\frac{\sinh\left( z p_0 /2 \right)}{z/2}\right)^2 - e^{{z} p_0} p_1^2.
\nonumber
\ee
Again, it should be noted that  these expressions are just~\eqref{kdsa11b},~\eqref{cop11mink} and~\eqref{kcas11b} under the identification~\eqref{id11}.

The parametrization of the orbit is now just 
\begin{eqnarray}
&& S_0 =\sinh(z p_0) \, +\frac{1}{2}\, z^2\,e^{z\,p_0}\, p_1^2, \nonumber\\
&& S_1 = z\,  e^{z\,p_0}\,p_1, \nonumber\\
&& S_2 = 0, \nonumber \\
&& S_3 = \cosh(z p_0) \, -\frac{1}{2}\, z^2\,e^{z\,p_0}\, p_1^2 , \nonumber
\nonumber
\end{eqnarray}  
so we can describe it as
\begin{align}
&\left\{\left(S_0,S_1,0,S_3 \right) \in \mathbb{R}^{1,3} : -S_0^2 + S_1^2 + S_3^2 =1,\ S_0+S_3>0\right\} \nonumber \\
&\qquad   =\left\{\left(S_0,S_1,S_3 \right) \in \mathbb{R}^{1,2} : -S_0^2 + S_1^2 + S_3^2 =1,\ S_0+S_3>0\right\},
\nonumber
\end{align}
where the first description is the intersection of~\eqref{ms11set} with the hyperplane $S_2=0$ and the second one is the description as half of a de Sitter spacetime in $(1+1)$ dimensions. This interpretation will be important in higher dimensional generalizations of this construction.

%%%%%%%%%%%%%%%%%%%%%%%%%%%%%%%%%%%%%%%%%%%%%%%%%%%%%%%%%%

\section{The $\kappa$-$G_\Lambda$ quantum group in (2+1) dimensions}\label{kappa21}

In this section we present the analogous of the previous construction to the case of $(2+1)$ dimensions. The main ideas of the construction are essentially unchanged, so the emphasis is concentrated in presenting the relevant expressions in such a way that the $(1+1)$-dimensional analogues are obtained by canonical Lie algebra projection $\pi_\mathfrak{g}^{2 \rightarrow 1} : \gc 2 \rightarrow \gc 1$. As it has been mentioned before, the (2+1)-dimensional case is specially interesting because of the outstanding role that quantum groups and their semiclassical counterparts, Poisson-Lie groups, play as the underlying symmetries for (2+1) gravity. Moreover, the topological nature of this theory implies that a unified consideration of the solutions for positive, zero and negative cosmological constant provides a unified description of all the possible solutions, at least locally.

%%%%%%%%%%%%%%%%%%%%%%%%%%%%%%%%%%%%%%%%%%%%%%%%%%%%%%%%
 
\subsection{Coboundary Lie bialgebra structure $\kgc 2$}
Commutation relations in a kinematical basis for $\gc 2$ are given by~\eqref{ds21}. The two quadratic Casimir functions for~\eqref{ds21} are 
\be
{\cal C}=P_0^2-\mathbf{P}^2- \Lambda(J^2-\mathbf{K}^2), \qquad
{\cal W}=-JP_0+K_1P_2-K_2P_1  ,
\label{cas3}
\ee
where $\mathbf{P}^2=P_1^2+P_2^2$ and $\mathbf{K}^2=K_1^2+K_2^2$. Recall that $\cal C$ comes from the Killing--Cartan form and is related to the energy of a point particle, while $\cal W$ is the Pauli--Lubanski vector. 

The coboundary Lie bialgebra $\kgc 2$ is defined by the following $r$-matrix:
\be
r^{2+1}_\c=z(K_1 \wedge P_1 + K_2 \wedge P_2)   .
\label{r21}
\ee
Hence the associated cocommutator map  is  given by
\begin{eqnarray}
&& \delta_\kappa(P_0) =  \delta_\kappa(J)=0 , \nonumber\\ 
&&  \delta_\kappa (P_1)=   z (P_1\wedge P_0  + \Lambda \, K_2\wedge J  )  ,
\nonumber\\ 
&&
  \delta_\kappa(P_2)=  z  (P_2\wedge P_0 - \Lambda\,  K_1 \wedge J)       ,
\label{cc}  \\
&&
  \delta_\kappa(K_1)= z (K_1 \wedge P_0  + P_2 \wedge J)   ,
 \nonumber\\ 
&&
  \delta_\kappa(K_2)=   z  (K_2  \wedge P_0 -P_1\wedge J)   .
  \nonumber
\end{eqnarray}

  %%%%%%%%%%%%%%%%%%%%%%%%%%%%%%%%%%%%%%%%%%%%%%%%%%%%%%%%%%

\subsection{Poisson-Hopf algebra structure on $\mathcal{U}_z(\gc 2)$}

The Poisson version of $\gc 2$  (\ref{ds21}) reads
\be
\begin{array}{lll} 
\pois{J}{P_i}=   \epsilon_{ij}P_j , &\qquad
\pois{J}{K_i}=   \epsilon_{ij}K_j , &\qquad  \pois{J}{P_0}= 0  , \\[2pt]
\pois{P_i}{K_j}=-\delta_{ij}P_0 ,&\qquad \pois{P_0}{K_i}=-P_i ,&\qquad
\pois{K_1}{K_2}= -J   , \\[2pt]
\pois{P_0}{P_i}=-\Lambda\, K_i ,&\qquad \pois{P_1}{P_2}= \Lambda\, J  ,
\end{array}
\nonumber
\ee
and the undeformed Poisson-Hopf structure on $\mathcal{U}(\gc 2)$ is given by the primitive coproduct 
$$
\Delta_0 (X)={X \otimes 1 + 1 \otimes X} , \qquad \forall X \in \gc 2 .
$$

We will use the bicrossproduct type basis~\cite{Majid:1994cy}. In this basis the Poisson-Hopf algebra structure $\mathcal{U}_z(\gc 2)$ is the Hopf algebra deformation with parameter $z=1/\kappa$ given by~\cite{ck3, ahep, starodutsev}
\be
\begin{array}{lll} 
  \pois{J}{P_1}=   P_2 ,
& \pois{J}{P_2}=  -P_1 ,   &\pois{J}{P_0}=   0    
  , \\[2pt]
\pois{J}{K_1}=     K_2    ,
&   \pois{J}{K_2}=     -K_1   ,
&   \pois{K_1}{K_2}=  - \displaystyle{ \frac{  \sin(2 z \sqrt{\Lambda} J)}{2z \sqrt{\Lambda} }  } , \\[8pt]
   \pois{P_0}{P_1}=  - \Lambda\,K_1 ,
&    \pois{P_0}{P_2}=   - \Lambda\,K_2  ,
&  \pois{P_1}{P_2}=   \Lambda\,  \displaystyle{ \frac{  \sin(2 z \sqrt{ \Lambda} J)}{2z \sqrt{\Lambda} }} , 
\end{array}
\nonumber
\ee
\vskip-0.75cm
\bea
&& \pois{K_1}{P_0}=     P_1  ,\qquad\qquad\qquad\qquad\;\;\;
    \pois{K_2}{P_0}=     P_2,\label{pa21}\\
&& \pois{P_2}{K_1}=     z \left(P_1 P_2 - \Lambda K_1 K_2\right)   , \qquad \pois{P_1}{K_2}=    z \left( P_1 P_2- \Lambda K_1 K_2  \right)      ,
\nonumber\\
&& \pois{K_1}{P_1}=      \frac{1}{2z} \left(  \cos(2z\sqrt{\Lambda} J) - e^{-2zP_0}    \right)  +\frac{z}{2} \left( P_2^2-P_1^2\right)  - \frac{z \Lambda}{2}  \left( K_2^2-K_1^2 \right)  ,\nonumber\\
&& \pois{K_2}{P_2}=   \frac{1}{2z} \left( \cos(2z \sqrt{ \Lambda}  J) - e^{-2zP_0} \right) +\frac{z}{2} \left(P_1^2 -P_2^2\right)-\frac{z\Lambda}{2} \left(  K_1^2 -K_2^2 \right)        ,\nonumber
\eea
and deformed coproduct map 
\bea
\Delta_\kappa ( P_0 ) \!\!\!&=&\!\!\!  P_0 \otimes 1+1 \otimes P_0   ,\qquad \Delta_\kappa  ( J  ) =  J \otimes 1 +1 \otimes J , \nonumber\\
\Delta_\kappa ( P_1 )  \!\!\!&=&\!\!\!  P_1\otimes   \cos (z\sqrt{ \Lambda} J) +e^{-z P_0}  \otimes P_1 
+  \Lambda\, K_2 \otimes    \frac{ \sin (z\sqrt{\Lambda} J) } {\sqrt{ \Lambda}}  , \nonumber \\
\Delta_\kappa ( P_2 ) \!\!\!&=&\!\!\!  P_2\otimes  \cos( z\sqrt{\Lambda} J) +e^{-z P_0}  \otimes P_2
- \Lambda\, K_1 \otimes   \frac{ \sin (z\sqrt{\Lambda} J) } {\sqrt{ \Lambda}}  , \label{cop21} \\
\Delta_\kappa ( K_1 ) \!\!\!&=&\!\!\!  K_1\otimes     \cos( z\sqrt{ \Lambda} J) +e^{-z P_0}  \otimes K_1 +  P_2 \otimes       \frac{ \sin (z\sqrt{\Lambda} J) } {\sqrt{\Lambda}}   ,  \nonumber \\
\Delta_\kappa ( K_2 ) \!\!\!&=&\!\!\!  K_2\otimes    \cos (z\sqrt{\Lambda} J) +e^{-z P_0}  \otimes K_2 -  P_1 \otimes      \frac{ \sin (z\sqrt{\Lambda} J) } {\sqrt{\Lambda}}  ,
\nonumber
\eea 
which explicitly depends on the cosmological constant $\c$ (compare with (\ref{cop11})).

The deformed counterpart of the Casimir function ${\cal C}$ (\ref{cas3})  for this Poisson-Hopf algebra reads
\be
{\cal C}_z=\frac 2{z^2}\left[ \cosh (zP_0)\cos(z\sqrt{ \Lambda} J)-1 \right]
-e^{zP_0} \left( \mathbf{P}^2 - \Lambda \,\mathbf{K}^2 \right)\,\cos(z\,\sqrt{\Lambda}\,J)
-2\, \Lambda \,e^{zP_0}\,\frac{\sin(z \sqrt{\Lambda}J)}{\sqrt{ \Lambda}}\,(K_1 P_2-K_2P_1) .
\nonumber
\ee
 When constructing the associated momentum space in the next section, we will see that a given projection of $\mathcal{C}_z$ will play the role of the deformed dispersion relation.

%%%%%%%%%%%%%%%%%%%%%%%%%%%%%%%%%%%%%%%%%%%%%%%%%%%%%%%%

\subsection{Non-commutative spacetime induced by a Poisson-Lie structure on $\Gc 2$}
As in the $(1+1)$-dimensional case, the fact that the $\kappa$-deformation is defined by a classical $r$-matrix allows us to construct a Poisson structure on $\Gc 2$ such that group multiplication is a Poisson homomorphism (see the Appendix for details on $\Gc 2$ and the expressions of left- and right-invariant vector fields which are presented in Table \ref{IVF21}). 

This canonical Poisson structure is given by the so-called Sklyanin bracket (\ref{skb}) on $\Gc 2$, now for the classical $r$-matrix (\ref{r21}). The Poisson subalgebra for spacetime coordinates can be interpreted as a non-commutative spacetime. The fundamental Poisson brackets are given by
\begin{align}
\pois{x^1}{x^0}&=z\; \frac{\tanh{(\sqrt{-\c} x^1)}}{\sqrt{-\c} \cosh^2{(\sqrt{-\c} x^2)}} = z\; \frac{\tan{(\sqrt{\c} x^1)}}{\sqrt{\c} \cos^2{(\sqrt{\c} x^2)}}, \nonumber \\
\pois{x^2}{x^0}&=z\; \frac{\tanh{(\sqrt{-\c} x^2)}}{\sqrt{-\c}} = z\; \frac{\tan{(\sqrt{\c} x^2)}}{\sqrt{\c}}, \nonumber \\
\pois{x^1}{x^2}&=0 .\nonumber
\end{align}
Quite remarkably, this Poisson non-commutative spacetime can be trivially quantized as no ordering problems arise due to commutativity of space coordinates. The full non-commutative spacetime is thus given by 
\begin{align}
\conm{\hat x^1}{\hat x^0}&=z\; \frac{\tanh{(\sqrt{-\c} \hat x^1)}}{\sqrt{-\c} \cosh^2{(\sqrt{-\c} \hat x^2)}} = z\; \frac{\tan{(\sqrt{\c} \hat x^1)}}{\sqrt{\c} \cos^2{(\sqrt{\c} \hat x^2)}}, \nonumber \\
\conm{\hat x^2}{\hat x^0}&=z\; \frac{\tanh{(\sqrt{-\c} \hat x^2)}}{\sqrt{-\c}} = z\; \frac{\tan{(\sqrt{\c} \hat x^2)}}{\sqrt{\c}}, \nonumber \\
\conm{\hat x^1}{\hat x^2}&=0,\nonumber
\end{align}
where $\{\hat x^0,\hat x^1,\hat x^2\}$ are the generators of the non-commutative algebra.

%%%%%%%%%%%%%%%%%%%%%%%%%%%%%%%%%%%%%%%%%%%%%%%%%%%%%%%%%

\subsection{Poincar\'e-Minkowski limit: contraction $\c \to 0$}

As we did in the $(1+1)$-dimensional case, we perform the limit $\c \to 0$ to recover the well-known expressions for the Poisson analogue of the $\kappa$-Poincar\'e algebra, which reads
\bea
&&   \pois{J}{P_1}=   P_2 ,
\qquad \pois{J}{P_2}=  -P_1 , \qquad \pois{J}{P_0}=   0     
  ,  \nonumber \\ 
&&  \pois{J}{K_1}=     K_2    ,\quad\ \,
   \pois{J}{K_2}=     -K_1   ,\quad\ \ 
    \pois{K_1}{K_2}= -  J,  \nonumber \\ 
&&   \pois{P_0}{P_1}=  0 ,\qquad
     \pois{P_0}{P_2}=   0  ,\qquad
   \pois{P_1}{P_2}=  0 , \nonumber \\ 
    && \pois{K_1}{P_0}=     P_1  ,\qquad\quad  \;\; 
    \pois{K_2}{P_0}=     P_2,\label{pa21mink}\\
&& \pois{P_2}{K_1}=     z  P_1 P_2  ,   \qquad \pois{P_1}{K_2}=    z P_1 P_2    ,
\nonumber\\
&& \pois{K_1}{P_1}=      \frac{1}{2z} \left( 1 - e^{-2zP_0}    \right)  +\frac{z}{2} \left( P_2^2-P_1^2\right)  ,\nonumber\\
&& \pois{K_2}{P_2}=   \frac{1}{2z} \left( 1 - e^{-2zP_0} \right) +\frac{z}{2} \left(P_1^2 -P_2^2\right)       ,\nonumber
\eea
and the deformed coproduct map is
\bea
\Delta_\kappa ( P_0 ) \!\!\!&=&\!\!\!  P_0 \otimes 1+1 \otimes P_0   ,\nonumber\\
\Delta_\kappa  ( J  ) \!\!\!&=&\!\!\!  J \otimes 1 +1 \otimes J , \nonumber\\
\Delta_\kappa ( P_1 )  \!\!\!&=&\!\!\!  P_1\otimes   1 +e^{-z P_0}  \otimes P_1  , \nonumber \\
\Delta_\kappa ( P_2 ) \!\!\!&=&\!\!\!  P_2\otimes  1+e^{-z P_0}  \otimes P_2 , \label{cop21mink} \\
\Delta_\kappa ( K_1 ) \!\!\!&=&\!\!\!  K_1\otimes   1 +e^{-z P_0}  \otimes K_1 + z P_2 \otimes  J  ,  \nonumber \\
\Delta_\kappa ( K_2 ) \!\!\!&=&\!\!\!  K_2\otimes 1 +e^{-z P_0}  \otimes K_2 - z P_1 \otimes J  .
\nonumber
\eea 
The deformed $\kappa$-Poincar\'e Casimir 
\bea
{\cal C}_z\!\!\!&=&\!\!\! \frac 2{z^2}\left[ \cosh (zP_0)-1 \right] \label{defcas21mink}
-e^{zP_0} \,(P_1^2+P_2^2)
\eea
is greatly simplified in this limit and in particular it does not depend on boost and rotation generators. 

Finally, the associated  non-commutative Poisson Minkowski spacetime is defined by the fundamental brackets
\begin{align}
\pois{x^1}{x^0}&=z\; x^1, \nonumber \\
\pois{x^2}{x^0}&=z\; x^2, \nonumber \\
\pois{x^1}{x^2}&=0 , \nonumber
\end{align}
and the full non-commutative spacetime is given by
\begin{align}
\conm{\hat x^1}{\hat x^0}&=z\; \hat x^1, \nonumber \\
\conm{\hat x^2}{\hat x^0}&=z\; \hat x^2, \nonumber \\
\conm{\hat x^1}{\hat x^2}&=0 , \nonumber
\end{align}
which is the $(2+1)$-dimensional version of the well-known $\kappa$-Minkowski spacetime~\cite{kMinkowski}.

%%%%%%%%%%%%%%%%%%%%%%%%%%%%%%%%%%%%%%%%%%%%%%%%%%%%%%%%%%

\section{Curved momentum spaces in (2+1) dimensions}\label{kappa21CMS}

In this section we describe the associated momentum space to the $\kappa$-deformation in $(2+1)$ dimensions. We carry out that by following the same procedure as in the $(1+1)$-dimensional case: we firstly construct the dual Lie algebra $(\gc 2)^\ast$ and afterwards we make use of  the quantum duality principle to obtain the dual quantum group. The momentum space will arise as an appropriate orbit of the latter acting on a suitable ambient space.

%%%%%%%%%%%%%%%%%%%%%%%%%%%%%%%%%%%%%%%%%%%%%%%%%%%%%%%%

\subsection{Dual Poisson-Lie group}
The skew symmmetrized first-order in $z$ of the coproduct~\eqref{cop21} is given by the cocommutator map (\ref{cc}).  Denoting by $\{X^0,X^1,X^2,L^1,L^2,R\}$  the  generators dual to, respectively, $\{P_0,P_1,P_2,K_1,K_2,J\}$, the Lie brackets defining the Lie algebra $(\gc 2)^\ast$ of the dual Poisson-Lie group $(\Gc 2)^\ast$ are
\be
\begin{array}{lll} 
[X^0, X^1]=-z \, X^1    ,
& \qquad [X^0, X^2]=-z \, X^2  , 
& \qquad [X^1, X^2]=0
  , \\[2pt]
[X^0, L^1]=-z\,  L^1    ,
&\qquad [X^0, L^2]=-z \, L^2  , 
&\qquad [L^1, L^2]=0  ,
 \\[2pt]
[R, X^2]=-z \, L^1    ,
 &\qquad[R, L^1]=z\,{\Lambda}\,  X^2  , 
&\qquad [L^1, X^2]=0   ,
 \\[2pt]
[R, X^1]=z \, L^2    ,
 &\qquad[R, L^2]=-z\,{\Lambda}\,  X^1  , 
 &\qquad[L^2, X^1]=0  ,
 \\[2pt]
[R, X^0]=0    ,
&\qquad [L^1, X^1]=0  , 
 &\qquad[L^2, X^2]=0  .
\end{array}
\nonumber
\ee
A faithful representation $\rho : (\gc 2)^\ast \rightarrow \text{End}(\mathbb{R}^6)$ of this Lie algebra for $\c\neq 0$ (as in the $(1+1)$-dimensional case here we present expressions for $\c >0$, but an analogous construction is valid in the case of negative cosmological constant and the final expressions are valid for both cases)  turns out to be
\be
\rho(X^0)= z \left( 
\begin{array}{cccccc}
0 & 0 & 0 & 0 & 0 & 1\\
0 & 0 & 0 & 0 & 0 & 0\\
0 & 0 & 0 & 0 & 0 & 0\\
0 & 0 & 0 & 0 & 0 & 0\\
0 & 0 & 0 & 0 & 0 & 0\\
1 & 0 & 0 & 0 & 0 & 0
\end{array}
\right) ,   \quad 
\rho(X^1)= z \left( 
\begin{array}{cccccc}
0 & 1 & 0 & 0 & 0 & 0\\
1 & 0 & 0 & 0 & 0 & 1\\
0 & 0 & 0 & 0 & 0 & 0\\
0 & 0 & 0 & 0 & 0 & 0\\
0 & 0 & 0 & 0 & 0 & 0\\
0 & -1 & 0 & 0 & 0 & 0
\end{array}
\right)  ,
\nonumber
\ee
\be
\rho(X^2)= z \left( 
\begin{array}{cccccc}
0 & 0 & 1 & 0 & 0 & 0\\
0 & 0 & 0 & 0 & 0 & 0\\
1 & 0 & 0 & 0 & 0 & 1\\
0 & 0 & 0 & 0 & 0 & 0\\
0 & 0 & 0 & 0 & 0 & 0\\
0 & 0 & -1 & 0 & 0 & 0
\end{array}
\right) ,
 \quad 
\rho(L^1)= z \sqrt{ \Lambda} \left( 
\begin{array}{cccccc}
0 & 0 & 0 & 1 & 0 & 0\\
0 & 0 & 0 & 0 & 0 & 0\\
0 & 0 & 0 & 0 & 0 & 0\\
1 & 0 & 0 & 0 & 0 & 1\\
0 & 0 & 0 & 0 & 0 & 0\\
0 & 0 & 0 & -1 & 0 & 0
\end{array}
\right)  ,
\nonumber
\ee
\be
\rho(L^2)= z \sqrt{\Lambda} \left( 
\begin{array}{cccccc}
0 & 0 & 0 & 0 &1 & 0\\
0 & 0 & 0 & 0 & 0 & 0\\
0 & 0 & 0 & 0 & 0 & 0\\
0 & 0 & 0 & 0 & 0 & 0\\
1 & 0 & 0 & 0 & 0 & 1\\
0 & 0 & 0 & 0 & -1 & 0
\end{array}
\right) ,
\quad
\rho(R)= z \sqrt{ \Lambda} \left( 
\begin{array}{cccccc}
0 & 0 & 0 & 0 & 0 & 0\\
0 & 0 & 0 & 0 &-1 & 0\\
0 & 0 & 0 & 1 & 0 & 0\\
0 & 0 & -1 & 0 & 0 & 0\\
0 & 1 & 0 & 0 & 0 & 0\\
0 & 0 & 0 & 0 & 0 & 0
\end{array}
\right).
\nonumber
\ee
If we denote as $\{p_0,p_1,p_2,\chi_1,\chi_2,\theta \}$ the local group coordinates which are dual  to $\{X^0,X^1, X^2, L^1,L^{2}, R \}$, correspondingly,  then a Lie group element $h$ on $(\Gc 2)^\ast$ can be written, provided it is sufficiently close to the identity, as
\be
h=\exp \left(\theta \rho(R)\right)\exp \left(p_1 \rho(X^1)\right)\exp \left(p_2 \rho(X^2) \right) \exp \left( \chi_1\rho(L^1) \right)
\exp \left( \chi_2\rho(L^2) \right) \exp \left(p_0 \rho(X^0) \right) .
\nonumber
\ee
Its explicit expression can be straightforwardly computed, although we omit it here for the sake of brevity. The group law for $(\Gc 2)^\ast$ can be directly derived and written as the following coproduct map for the six group coordinates:
\begin{eqnarray}
&& \Delta(p_0) = p_0 \otimes 1 + 1 \otimes p_0, \qquad
\Delta(\theta)=\theta \otimes 1 + 1 \otimes \theta, \nonumber\\
&&\Delta(p_1)= p_1\otimes \cos(z\,\sqrt{\Lambda}\, \theta) +  e^{-{z} p_0} \otimes p_1 +\Lambda\,\chi_2\otimes 
\frac{\sin(z\, \sqrt{\Lambda}\, \theta)}{\sqrt{\Lambda}},
\nonumber\\
&&\Delta(p_2)= p_2\otimes \cos(z\, \sqrt{ \Lambda}\, \theta) +  e^{-{z} p_0} \otimes p_2  - \Lambda\,\chi_1\otimes
\frac{\sin(z\, \sqrt{\Lambda}\, \theta)}{\sqrt{\Lambda}}, \label{codual21}\\
&&\Delta(\chi_1)= \chi_1\otimes \cos(z\, \sqrt{ \Lambda}\, \theta) +  e^{-{z} p_0} \otimes \chi_1  + p_2\otimes 
\frac{\sin(z\, \sqrt{\Lambda}\, \theta)}{\sqrt{\Lambda}},\nonumber\\
&&\Delta(\chi_2)= \chi_2\otimes \cos(z\, \sqrt{ \Lambda}\, \theta) +  e^{-{z} p_0} \otimes \chi_2  - p_1\otimes \frac{\sin(z\, \sqrt{\Lambda}\, \theta)}{\sqrt{\Lambda}}.
\nonumber
\end{eqnarray}  
Again, under the identification
\be
p_0\equiv P_0, \quad
p_1\equiv P_1, \quad
p_2\equiv P_2, \quad
\chi_1\equiv K_1, \quad
\chi_2\equiv K_2, \quad
\theta \equiv J,
\label{id21}
\ee
this is exactly the coproduct for the Poisson-Hopf algebra $\mathcal{U}_z(\gc 2)$ given in~\eqref{cop21}, and the unique Poisson-Lie structure on $(\Gc 2)^\ast$ that is compatible with~\eqref{codual21} and has the undeformed Lie algebra $\gc 2$~\eqref{ds21} as its linearization is the deformed Poisson algebra given by~\eqref{pa21}.

%%%%%%%%%%%%%%%%%%%%%%%%%%%%%%%%%%%%%%%%%%%%%%%%%%%%%%%%

\subsection{Momentum space as an orbit and deformed dispersion relation}

As in the $(1+1)$-dimensional case, we consider the action $(\Gc 2)^\ast \triangleright \mathbb{R}^{1,5}$ given by left multiplication, the point $(0,0,0,0,0,1) \in \mathbb{R}^{1,5}$ and the orbit of $(\Gc 2)^\ast$ passing through it. This orbit is given in terms of Cartesian coordinates $S_i$ on $\mathbb{R}^{1,5}$ as the set
\be
\left\{\left(S_0,S_1,S_2,S_3,S_4,S_5 \right) \in \mathbb{R}^{1,5} : -S_0^2 + S_1^2 + S_2^2 + S_3^2 + S_4^2 + S_5^2 =1,\ S_0+S_5>0\right\},
\label{ms21setmink}
\ee
which is nothing but a subset of de Sitter spacetime in $(4+1)$ dimensions written as the usual embedding in $\mathbb{R}^{1,5}$. Moreover, we can use previous coordinates $\{p_0,p_1,p_2,\chi_1,\chi_2,\theta \}$ on the Lie group $(\Gc 2)^\ast$ to parametrize this space. Explicitly  
\begin{eqnarray}
&& S_0=  \sinh(z p_0) \, +\frac{1}{2}\,z^2\,e^{z\,p_0} \left[p_1^2 + p_2^2 +\Lambda \left(\chi_1^2+ \chi_2^2\right)\right], \nonumber\\
&& S_1= z\, e^{z\,p_0} \left(\cos(z\,\sqrt{\Lambda}\,\theta)\, p_1- \sqrt{\Lambda}\,\sin(z\,\sqrt{\Lambda}\,\theta)\,\chi_2 \right), \nonumber\\
&& S_2= z\, e^{z\,p_0} \left(\cos(z\,\sqrt{\Lambda}\,\theta)\, p_2 + \sqrt{\Lambda}\, \sin(z\,\sqrt{\Lambda}\,\theta)\,\chi_1 \right), \nonumber\\
&& S_3= z\, e^{z\,p_0}\left(    \sqrt{\Lambda}\, \cos(z\,\sqrt{\Lambda}\,\theta)\,\chi_1  -\sin(z\,\sqrt{\Lambda}\,\theta)\, p_2 \right ), \nonumber\\
&& S_4= z\, e^{z\,p_0}\left( \sqrt{\Lambda}\,\cos(z\,\sqrt{\Lambda}\,\theta)\,\chi_2  + \sin(z\,\sqrt{\Lambda}\,\theta)\, p_1\right), \nonumber\\
&& S_5=\cosh(z p_0) \, -\frac{1}{2}\,z^2\,e^{z\,p_0} \left[p_1^2 + p_2^2 +\Lambda \left(\chi_1^2+ \chi_2^2\right)\right]\,.\nonumber\end{eqnarray}    
Additionally, the condition 
\be
S_0+S_5=e^{z\,p_0}>0
\nonumber
\ee
makes clear that this orbit is half of a de Sitter spacetime in $(4+1)$ dimensions.

%%%%%%%%%%%%%%%%%%%%%%%%%%%%%%%%%%%%%%%%%%%%%%%%%%%%%%%%

\subsection{Poincar\'e-Minkowski $\c \to 0$ limit in momentum space}

Similarly to  the $(1+1)$-dimensional case, we finish by writing explicitly the Poisson-Hopf structure on the dual group $(\Gc 2)^\ast$ in the limit $\c \to 0$, which gives rise to the fundamental Poisson brackets given by
\bea
&&   \pois{\theta}{p_1}=   p_2 ,\qquad 
  \pois{\theta}{p_2}=  -p_1 , \qquad \pois{\theta}{p_0}=   0     
  ,\nonumber\\
&&   \pois{\theta}{\chi_1}=     \chi_2    ,\qquad\!
    \pois{\theta}{\chi_2}=     -\chi_1   ,
\quad\ \    \pois{\chi_1}{\chi_2}=  -  \theta  ,\nonumber\\
&&      \pois{p_0}{p_1}=  0 ,\qquad
    \pois{p_0}{p_2}=   0  ,\qquad
  \pois{p_1}{p_2}=  0 , 
\nonumber\\
&& \pois{\chi_1}{p_0}=     p_1  ,\quad\qquad\;\;\;
    \pois{\chi_2}{p_0}=     p_2,\nonumber\\
&& \pois{p_2}{\chi_1}=     z \; p_1 p_2   , \qquad \pois{p_1}{\chi_2}=    z \; p_1 p_2    ,
\nonumber\\
&& \pois{\chi_1}{p_1}=      \frac{1}{2z} \left( 1 - e^{-2z p_0}    \right)  +\frac{z}{2} \left( p_2^2-p_1^2\right)  ,\nonumber\\
&& \pois{\chi_2}{p_2}=   \frac{1}{2z} \left( 1 - e^{-2z p_0} \right) +\frac{z}{2} \left(p_1^2 -p_2^2\right)       ,\nonumber
\eea
along with  the following coproduct  
\begin{eqnarray}
&& \Delta(p_0) = p_0 \otimes 1 + 1 \otimes p_0, \nonumber\\
&& \Delta(\theta)=\theta \otimes 1 + 1 \otimes \theta, \nonumber\\
&& \Delta(p_1) =  p_1\otimes 1 +  e^{-{z} p_0} \otimes p_1 ,\qquad 
\nonumber\\
&& \Delta(p_2) =  p_2\otimes 1 +  e^{-{z} p_0} \otimes p_2 ,\nonumber\\
&& \Delta(\chi_1) =  \chi_1\otimes 1 +  e^{-{z} p_0} \otimes \chi_1  + z\; p_2\otimes \theta,\nonumber\\
&& \Delta(\chi_2) =  \chi_2\otimes 1 +  e^{-{z} p_0} \otimes \chi_2  - z\; p_1\otimes \theta.
\nonumber
\end{eqnarray}
The Casimir function for this Poisson structure reads
\be
{\cal C}_z= \frac 2{z^2}\left[ \cosh (z p_0)-1 \right]-e^{z p_0} \,(p_1^2+p_2^2)\,.
\nonumber
\ee
It should be noted that, again, these expressions are just~\eqref{pa21mink},~\eqref{cop21mink} and~\eqref{defcas21mink} under the identification~\eqref{id21}. 

Now the parametrization of the orbit is just
\begin{eqnarray}
&& S_0=  \sinh(z p_0) \, +\frac{1}{2}\,z^2\,e^{z\,p_0} \left(p_1^2 + p_2^2 \right), \nonumber\\
&& S_1= z\, e^{z\,p_0}\, p_1, \nonumber\\
&& S_2= z\, e^{z\,p_0}\,p_2, \nonumber\\
&& S_3= 0, \nonumber\\
&& S_4= 0, \nonumber\\
&& S_5=\cosh(z p_0) \, -\frac{1}{2}\,z^2\,e^{z\,p_0} \left(p_1^2 + p_2^2 \right),\nonumber\end{eqnarray} 
so we can describe it as
\begin{align}
&\left\{\left(S_0,S_1,S_2,0,0,S_5 \right) \in \mathbb{R}^{1,5} : -S_0^2 + S_1^2 + S_2^2 + S_5^2 =1, \ S_0+S_5>0\right\}  \nonumber \\
&\qquad =\left\{\left(S_0,S_1,S_2,S_5 \right) \in \mathbb{R}^{1,3} : -S_0^2 + S_1^2 + S_2^2 + S_5^2 =1,\  S_0+S_5>0\right\},
\nonumber
\end{align}
where the first description corresponds to the intersection of~\eqref{ms21setmink} with the hyperplane $S_3=S_4=0$ and the second one is the description as half of a de Sitter spacetime in $(2+1)$ dimensions, as it was originally obtained in~\cite{Kowalski-Glikman:2013rxa}.

%%%%%%%%%%%%%%%%%%%%%%%%%%%%%%%%%%%%%%%%%%%%%%%%%%%%%%%%%%

\section{Concluding remarks}\label{kappa31}

We have presented a review of recent developments relating non-commutative spacetimes and curved momentum spaces arising from the so-called $\kappa$-(anti-)de Sitter quantum deformation, which is the analogous of the well known $\kappa$-Poincar\'e deformation with non-vanishing cosmological constant. Our construction treats in a unified way all the $\kappa$-deformations, regardless of the value of the cosmological constant, and thus allows for an easier interpretation of the global consequences of the quantum deformation. In particular, all the expressions are continuous with respect to the cosmological constant parameter $\c$, producing in this way a uniform description of the Poincar\'e, de Sitter and anti-de Sitter cases. 

While we focussed on $(1+1)$ and $(2+1)$ dimensional models, both the quantum spacetimes and the curved momentum spaces constructed in this paper can be generalized to the physically relevant $(3+1)$-dimensional case, as shown in~\cite{BGGH31}. Again, the cases with positive and negative cosmological constant  give rise to different momentum spaces: when $\c > 0$ (de Sitter spacetime) the momentum space is half of a $(6+1)$-dimensional de Sitter spacetime, while when $\c < 0$ (anti-de Sitter spacetime) the momentum space can be identified with half of an $SO(4,4)$-quadric. 

 As it is discussed in~\cite{BGGH31} and rigorously studied in \cite{BGH31}, the essential structural difference that arises in the $(3+1)$-dimensional case is related with the appearance of the term $z\sqrt{-\c} \,J_1 \wedge J_2$ in the classical $r$-matrix~\eqref{r31ads}, which is not present either in lower dimensional cases or in the case of vanishing cosmological constant~\eqref{rkappamink}. Again, a remarkable property of both the $(6+1)$-dimensional de Sitter spacetime and the $SO(4,4)$-quadric is that a $(3+1)$-dimensional de Sitter spacetime is obtained as an intersection with some hyperplane, giving in this way a clear geometrical interpretation of the vanishing cosmological constant limit.

We conclude by providing an outlook on a few open issues that the results presented here will allow to understand better. As mentioned in the preceding paragraph, our improved understanding of the new structures that emerge in going from lower-dimensional cases to the (3+1) model allows to construct  explicitly  the (3+1) non-commutative spacetime corresponding to the $\kappa$-deformation in presence of a non-vanishing cosmological constant $\c$. This is already work in progress and will be presented in~\cite{BGH31}. Also, the very same techniques developed here can be applied  to the  construction of the curved momentum spaces associated to other quantum deformations. In the same spirit, the study of non-relativistic limits of the construction here described is work in progress and will be given elsewhere.  Finally, the work we presented here could be generalised to describe the momentum space associated to deformed symmetries of spacetime models that are not maximally symmetric. For example, it would be of particular interest to analyse the momentum space associated to a quantum-deformed Friedmann-Lemaître-Robertson-Walker (FLRW) geometry, which is relevant from a cosmological point of view. 
The case of a homogeneous and isotropic spacetime in the framework of the $\kappa$-deformation was considered from a phenomenological point of view in \cite{BBGLP2017}, where general dispersion relations for a particle moving on a  FLRW geometry were derived.
Finding a way to perform the symmetry breaking under which the static (A)dS results here presented could be connected to a FLRW-type model is indeed worthy of being investigated. A possible line of attack would rely on a procedure that was employed in \cite{Marciano:2010gq, Rosati:2015pga} to describe relativistically-compatible deformations of particles worldlines propagating over such a quantum FLRW geometry. The fundamental ingredient that allowed to identify such worldlines relied on describing the FLRW spacetime as a series of 'slices' of de Sitter spacetime with different cosmological constants.

%%%%%%%%%%%%%%%%% Acknowledgments %%%%%%%%%%%%%%%%%%

\section*{Acknowledgments}

A.B., I.G-S. and F.J.H. have been partially supported by the grant MTM2016-79639-P (AEI/FEDER, UE), by Junta de Castilla y Le\'on (Spain) under grant VA057U16 and by the Action MP1405 QSPACE from the European Cooperation in Science and Technology (COST). I.G-S. acknowledges a PhD grant from the Junta de Castilla y Le\'on (Spain) and European Social Fund. G.G. acknowledges a Grant for Visiting Researchers at the Campus of International Excellence ``Triangular-E3" (MECD, Spain).

%%%%%%%%%%%%%%%%%%%%%%%%%%%%%%%%%%%%%%%%%%%%%%%%%%%%%%%%%%

\section*{Appendix: Invariant vector fields}

\setcounter{equation}{0}
\renewcommand{\theequation}{A.\arabic{equation}}

In this appendix we will summarize useful expressions regarding $\Gc n$. Consider a Lie group $G$ with Lie algebra $\mathfrak{g}=\text{Lie}\; G$ and fix a basis $\{X_i\}$ of $\mathfrak{g}$. Then
$$
[X_i,X_j]=c_{ij}^{k} \; X_k, \;\;\;\; i,j,k \in \{ 1,\ldots, N\} ,
$$
where $c_{ij}^{k}$ are the structure constants with respect to this basis and $N=\text{dim} \; \mathfrak{g}$. From a faithful representation $\rho : \mathfrak{g} \rightarrow \text{End}(V)$ of the Lie algebra $\mathfrak{g}$ on some vector space $V$,  we can always introduce local coordinates on the Lie group using the so called exponential coordinates of the second kind $\{\alpha^i \}$, in terms of which an element of the group is written as 
$$
g=\prod_{i=1}^N \exp{\left(\alpha^i \rho(X_i) \right)} .
$$

We denote by $X^L_i, X^R_i$ left- and right-invariant vector fields, respectively, defined by their action on functions $f \in \mathcal{C}^\infty (G)$:
\begin{align}
X^L_i f(g)&=\frac{{\rm d}}{{\rm d}t}\biggr\rvert _{t=0} f\left(g \, e^{t X_i}\right), & X^R_i f(g)=\frac{{\rm d}}{{\rm d}t}\biggr\rvert _{t=0} f\left(e^{t X_i} g\right) .
\nonumber
\end{align}

The Lie brackets of these vector fields satisfy 
\begin{align}
[X^L_i,X^L_j]&=c_{ij}^{k}  \; X^L_k, & [X^R_i,X^R_j]&=-c_{ij}^{k}  \; X^R_k\, .
\nonumber
\end{align}

We start with the $(1+1)$-dimensional case and consider the faithful representation for the Lie algebra $\rho : \gc 1 \rightarrow \text{End}(\mathbb{R}^3)$     given by
\bea
\rho(P_0)=\begin{pmatrix}
0 & \c & 0\\ 
1 & 0 & 0\\
0 & 0 & 0\\
\end{pmatrix}, \qquad
\rho(P_1)=\begin{pmatrix}
0 & 0 & -\c\\ 
0 & 0 & 0\\
1 & 0 & 0\\
\end{pmatrix}, \qquad
\rho(K)=\begin{pmatrix}
0 & 0 & 0\\ 
0 & 0 & 1\\
0 & 1 & 0\\
\end{pmatrix}, \nonumber
\eea
and we introduce local coordinates on $\Gc 1$ using exponential coordinates of the second kind given by 
\be
g=\exp \left( x^0 \rho(P_0) \right) \exp \left( x^1 \rho(P_1) \right) \exp \left( \xi \; \rho(K) \right) \,.
\nonumber
\ee
Note that this coordinates $\{x^0, x^1, \xi \equiv \xi^1\}$ have a direct physical interpretation as time, space and boost (rapidity) coordinates, respectively. Left- and right-invariant vector fields in this coordinates are written in Table \ref{IVF11} in terms of a parameter $\ro = \sqrt{-\c}$.

%%%%%%%%%%%%%%%%%%%%%%%%%%%%%%%%%%%%%%%

\begin{table}[t]
{\footnotesize
 \noindent
\caption{{Left- and right-invariant  vector fields on the isometry groups of  the (1+1)-dimensional  de Sitter ($\Lambda>0$),  anti-de Sitter ($\Lambda<0$) and Minkowski ($\Lambda=0$) spaces in terms of $\ro = \sqrt{-\c}$.}}
\label{IVF11}
$$
\begin{array}{l} 
\hline
 \\[-4pt]
\displaystyle{\  X^L _{P_0} = \frac{1}{\cosh(\ro x^1)} \left( \cosh \xi \, \partial_{x^0} + \cosh (\ro x^1) \sinh \xi \, \partial_{x^1}  - \ro \sinh(\ro x^1) \cosh \xi \, \partial_\xi \right) \ } \\[12pt]
\displaystyle{\  X^L _{P_1} = \frac{1}{\cosh(\ro x^1)} \left( \sinh \xi \, \partial_{x^0} + \cosh (\ro x^1) \cosh \xi \, \partial_{x^1}  - \ro \sinh(\ro x^1) \sinh \xi \, \partial_\xi  \right) \ } \\[12pt]
\displaystyle{\  X^L _K = \partial _\xi \ } \\[12pt]
\hline
 \\[-4pt]
\displaystyle{\  X^R _{P_0} = \partial _{x^0}  \quad} \\[10pt]
\displaystyle{\  X^R _{P_1} = \frac{1}{\cosh(\ro x^1)} \left(- \sin(\ro x^0)  \sinh(\ro x^1)\, \partial_{x^0} + \cos(\ro x^0)\cosh(\ro x^1) \, \partial_{x^1}  - \ro \sin(\ro x^0) \,\partial_\xi \right)  \quad} \\[10pt]
\displaystyle{\ X^R _K = \frac{1}{\cosh(\ro x^1)} \left( \frac{  \cos(\ro x^0) \sinh(\ro x^1)}{\ro}\, \partial_{x^0} + \frac{ \sin(\ro x^0)\cosh(\ro x^1)}{\ro} \,\partial_{x^1}+ \cos(\ro x^0)\, \partial_\xi  \right) \quad} \\[10pt]
\hline
\end{array}
$$
}
\end{table}

Following the very same previous procedure, we consider the  $(2+1)$-dimensional case  and   begin with a faithful representation for the Lie algebra $\rho : \gc 2 \rightarrow \text{End}(\mathbb{R}^4)$, which takes the explicit form
\be
\begin{array}{l} 
\rho(P_0)=\left(\begin{array}{cccc}
0&\c&0&0\cr 
1&0&0&0\cr 
0&0&0&0\cr 
0&0&0&0
\end{array}\right) , \quad\ 
\rho(P_1)=\left(\begin{array}{cccc}
0&0&-\c&0\cr 
0&0&0&0\cr 
1&0&0&0\cr 
0&0&0&0
\end{array}\right) , \quad\ 
\rho(P_2)=\left(\begin{array}{cccc}
0&0&0&-\c\cr 
0&0&0&0\cr 
0&0&0&0\cr 
1&0&0&0
\end{array}\right) , \\[25pt]
\rho(J)=\left(\begin{array}{cccc}
0&0&0&0\cr 
0&0&0&0\cr 
0&0&0&-1\cr 
0&0&1&0
\end{array}\right)  , \quad\ 
\rho(K_1)=\left(\begin{array}{cccc}
0&0&0&0\cr 
0&0&1&0\cr 
0&1&0&0\cr 
0&0&0&0
\end{array}\right) , \quad\ 
\rho(K_2)=\left(\begin{array}{cccc}
0&0&0&0\cr 
0&0&0&1\cr 
0&0&0&0\cr 
0&1&0&0
\end{array}\right)  .
\end{array}
\nonumber
\ee
Next we introduce local coordinates on $\Gc 2$ using exponential coordinates of the second kind in the form
\be
g=\exp(x^0 \rho(P_0))\exp(x^1 \rho(P_1))\exp(x^2 \rho(P_2)) \exp(\xi^1 \rho(K_1))\exp(\xi^2 \rho(K_2))
\exp(\theta J)  .
\nonumber
\ee
This coordinates $\{x^0,x^1,x^2,\xi^1,\xi^2,\theta\}$ have a direct physical interpretation as time, space,  boosts  and rotation coordinates, respectively. In Table \ref{IVF21} left- and right-invariant vector fields are given in terms of these coordinates.

 %%%%%%%%%%%%%%%%%%%%%%%%%%%%%%%%%%%%%%%

\begin{table}[hbp] 
{\footnotesize  
 \noindent
\caption{{Left- and right-invariant  vector fields on the isometry groups of  the (2+1)-dimensional de Sitter ($\Lambda>0$),  anti-de Sitter ($\Lambda<0$) and Minkowski ($\Lambda=0$) spaces in terms of $\ro = \sqrt{-\c}$.}}
\label{IVF21}
$$
\begin{array}{l} 
\hline
 \\[-4pt]
 \displaystyle{\  X_{P_0}^L=\frac{\cosh\xi^1\cosh\xi^2}{\cosh( \ro x^1)\cosh( \ro
x^2)}\, \left( \partial_{x^0} -\ro \sinh( \ro x^1)\partial_{\xi^1} \right)
+\frac{\sinh\xi^1\cosh\xi^2}{ \cosh
(\ro x^2)}\,\partial_{x^1}+\sinh\xi^2\,\partial_{x^2}  -\ro\tanh (\ro
x^2) \cosh\xi^2\,\partial_{\xi^2} \ } \\[12pt]
\displaystyle{\   X_{P_1}^L=\left(
\frac{ \sinh\xi^1 \cos\theta  +\cosh\xi^1\sinh\xi^2 \sin\theta}{\cosh ( \ro
x^1)\cosh
(\ro x^2)}\right)\left( \partial_{x^0} -\ro \sinh( \ro x^1)\partial_{\xi^1} \right) +\left(
\frac{ \cosh\xi^1\cos\theta  +\sinh\xi^1\sinh\xi^2\sin\theta}{ \cosh
(\ro x^2)}\right)\partial_{x^1} \  }\\[8pt]
\displaystyle{\qquad\qquad +\cosh \xi^2\sin\theta \,\partial_{x^2}
-\ro \tanh( \ro x^2)   \left(    \tanh\xi^2  \cos\theta
\,\partial_{\xi^1}+  \sinh\xi^2 \sin\theta
\,\partial_{\xi^2}-   \frac{\cos\theta    }{\cosh\xi^2 } 
\,\partial_{\theta}      \right) \ }\\[12pt]
\displaystyle{\   X_{P_2}^L=\left(
\frac{ \cosh\xi^1\sinh\xi^2\cos\theta  - \sinh\xi^1\sin\theta}{\cosh( \ro
x^1)\cosh
(\ro x^2)}\right) \left( \partial_{x^0} -\ro \sinh( \ro x^1)\partial_{\xi^1} \right) +\left(
\frac{\sinh\xi^1\sinh\xi^2\cos\theta  - \cosh\xi^1 \sin\theta}{ \cosh
(\ro x^2)}\right)\partial_{x^1} \ }\\[8pt]
\displaystyle{\qquad\qquad +\cosh \xi^2\cos\theta\,\partial_{x^2}
+\ro  \tanh( \ro x^2)  \left(  \tanh\xi^2 \sin\theta
\,\partial_{\xi^1}-  \sinh\xi^2    \cos\theta
\,\partial_{\xi^2}-   \,\frac{\sin\theta }{\cosh\xi^2 } 
\,\partial_{\theta}  \right) \quad}\\[12pt]
\displaystyle{\   X_{K_1}^L= 
\frac{\cos\theta  }{\cosh  
\xi^2} \,\partial_{\xi^1} +\sin\theta\,\partial_{\xi^2}+  \tanh  
\xi^2\cos\theta\,\partial_{\theta} \quad}\\[12pt]
\displaystyle{\   X_{K_2}^L= -
\frac{\sin\theta  }{\cosh  
\xi^2} \,\partial_{\xi^1} +\cos\theta\,\partial_{\xi^2}-  \tanh  
\xi^2\sin\theta \,\partial_{\theta} \quad}\\[12pt]
\displaystyle{\     X_{J}^L= \partial_{\theta}\quad}\\[8pt]
\hline
 \\[-4pt]
\displaystyle{\  X_{P_0}^R= \partial_{x^0}  \quad} \\[10pt]
\displaystyle{\   X_{P_1}^R= 
-\sin( \ro x^0)\tanh( \ro x^1) \, \partial_{x^0} 
+\cos( \ro x^0)\, \partial_{x^1} -\ro\,\frac{\sin( \ro x^0)}{\cosh( \ro
x^1)}\, \partial_{\xi^1} \quad}\\[12pt] 
\displaystyle{\   X_{P_2}^R=
-\frac{\sin( \ro x^0)\tanh( \ro x^2) }{\cosh( \ro x^1)}\left(   \partial_{x^0} - \ro  \sinh( \ro x^1) \,
 \partial_{\xi^1} \right)
-\cos( \ro x^0)\sinh( \ro x^1)\tanh( \ro x^2)\,\partial_{x^1}+
\cos( \ro x^0)\cosh( \ro x^1)\,\partial_{x^2} \quad}\\[8pt]
\displaystyle{\qquad\qquad +  {\ro} \left( \frac{ \cos( \ro x^0)\sinh( \ro
x^1)\sinh  \xi^1- \sin( \ro x^0) \cosh  \xi^1}{\cosh( \ro x^2)} \right) 
 \partial_{\xi^2} \quad}\\[8pt]
\displaystyle{\qquad\qquad +   {\ro}\left( \frac
{\cos( \ro x^0)\sinh( \ro x^1)\cosh  \xi^1- \sin( \ro x^0) \sinh  \xi^1} {\cosh( \ro x^2)
\cosh  \xi^2}\right) 
\left( \partial_{\theta} -  \sinh\xi^2 \partial_{\xi^1} \right)\quad }\\[12pt]
\displaystyle{\   X_{K_1}^R= 
\frac{\cos( \ro x^0)  \tanh( \ro x^1)}{\ro}\,\partial_{x^0} +\frac{\sin (\ro
x^0)}{\ro}\,\partial_{x^1} +\frac{\cos( \ro x^0)}{\cosh( \ro x^1)}\,
\partial_{\xi^1}\quad  }\\[12pt]
\displaystyle{ \   X_{K_2}^R= 
\frac{\cos( \ro x^0)  \tanh( \ro x^2)}{\ro \cosh( \ro x^1)}\left(  \partial_{x^0}- \ro \sinh( \ro x^1) \,
 \partial_{\xi^1}  \right)
-\frac{\sin( \ro x^0)  \sinh( \ro x^1)  \tanh( \ro x^2)}{\ro}\,\partial_{x^1}
+\frac{\sin( \ro x^0) \cosh( \ro x^1)}{\ro}\,\partial_{x^2}\quad  }
\\[8pt]
\displaystyle{\qquad\qquad + \left( \frac{ \sin( \ro x^0)  \sinh(\ro  x^1)
\sinh  \xi^1  + \cos( \ro x^0)\cosh  
\xi^1 }{\cosh( \ro x^2)}  \right)  \partial_{\xi^2} \quad}\\[8pt]
\displaystyle{\qquad\qquad +   \left( \frac
{\sin( \ro x^0)\sinh( \ro x^1)\cosh  \xi^1+ \cos( \ro x^0) \sinh  \xi^1} { \cosh( \ro x^2)
\cosh  \xi^2  }  \right) 
\left( \partial_{\theta} - \sinh\xi^2 \partial_{\xi^1}
\right)\quad }\\[12pt]
\displaystyle{\    X_{J}^R= 
-\frac{\cosh( \ro x^1)  \tanh( \ro x^2)}{\ro  }\,\partial_{x^1}
+\frac{  \sinh( \ro x^1) }{\ro}\,\partial_{x^2}
- \frac{\cosh( \ro x^1) }{\cosh( \ro
x^2)}\left(\cosh\xi^1 \tanh\xi^2 \, \partial_{\xi^1}  - \sinh\xi^1  \,
\partial_{\xi^2}  -  \frac{\cosh\xi^1  }{\cosh\xi^2 }\,\partial_{\theta} \right)
\quad}\\[10pt]
\hline
\end{array}
$$
}
\end{table}
%%%%%%%%%%%%%%%%%%%%%%%%%%%%%%%%%%%%%%%%%%%%%%%%%%%%%%%%%%

\newpage

%%%%%%%%%%%%%%%%%%%%%%%%%%%%%%%%%%%%%%%%%%%%%%%%%%%%%

 \end{document}